\documentclass[twocolumn]{aastex631}

\usepackage[utf8]{inputenc}
\usepackage{natbib}
\usepackage{graphicx}
\usepackage{hyperref}
\usepackage{multirow}
\usepackage{threeparttable}
\usepackage{array}  
\usepackage{booktabs}

\begin{document}

\title{Transition Zone from ExoEarth to ExoVenus: Critical Insight from Planets on the Edge}

\author[0000-0002-0436-1802]{Lisa Kaltenegger}
\affiliation{Carl Sagan Institute, Ithaca NY, 14853, United States}
\affiliation{Cornell University, Department of Astronomy, Ithaca NY, 14853, United States}
\email{lk433@cornell.edu}  

\author[0009-0006-7156-3152]{Abigail Bohl}
\affiliation{Carl Sagan Institute, Ithaca NY, 14853, United States}
\affiliation{Cornell University, Department of Astronomy, Ithaca NY, 14853, United States}

\author[0000-0000-0000-0000]{Connor Novak}
\affiliation{Department of Astronomy and Carl Sagan Institute, Cornell University, 122 Sciences Drive, Ithaca, NY 14853, USA}

\author[0009-0007-1562-2944]{Gillis Lowry}
\affiliation{Department of Astronomy and Carl Sagan Institute, Cornell University, 122 Sciences Drive, Ithaca, NY 14853, USA}
\affiliation{Department of Physics \& Astronomy, San Francisco State University, 1600 Holloway Ave., San Francisco, CA 94132, USA}

\begin{abstract}
Venus and Earth, planets with similar masses and radii, provide astonishingly different environments, raising the question of why these two rocky worlds evolved so differently. Explaining this divergence is a central challenge in comparative planetology and astrobiology.
Now, atmospheric studies of exoplanets can explore the possibilities of terrestrial planet evolution, especially in the era of JWST; however, we have not yet realized the critical insights rocky exoplanets in the Transition Zone (TZ) can provide. 
These TZ planets in the overlap between the Habitable Zone and Venus Zone can test how incident stellar irradiation affects surface water retention and habitability. 

Observations of exoplanets in the TZ can probe not only the emergence and decline of exoplanet habitability but also test hypotheses about Venus’ past, complementing upcoming Venus flagship missions.
Here, we analyze NASA's Exoplanet Archive, updated with \textit{Gaia} DR3 data, and provide a target list of 23 rocky exoplanets in the Earth-to-Venus Transition Zone, along with 2 planets in the Venus Evolution Zone (EV). We examine their demographics and quantify their potential for follow-up observations. 

This list of rocky TZ and EV planets allows observers to optimize target selection and observing strategy for specific instruments.
Characterizing exoplanets in the Transition Zone will be critical to investigating open questions about the emergence and decline of habitability on rocky worlds, the evolution from an exo-Earth to an exo-Venus, and Earth’s future. 

\end{abstract}

\section{Introduction}

To understand the emergence and decline of habitability on rocky worlds, it is critical to grasp the transition of a rocky planet from a water-rich, Earth-like world to a dry, Venus-like world. Venus and Earth offer very different views of rocky planets, raising questions about why these two worlds evolved so differently and providing evidence that rocky planets can evolve into dramatically different states. 
Exoplanets receiving stellar irradiation between Earth's and Venus's provide a framework for addressing this challenge and exploring the conditions under which Earth-like habitability emerges, persists, or disappears.

Among the more than 6,000 detected exoplanets to date \citep{NASA}, 45 rocky planets are currently known to orbit in the empirical Habitable Zone (HZ) \citep{bohl2026} and are prime targets for observations in the search for life in the cosmos. However, to explore the open question of habitability on rocky worlds in the region between Earth and Venus and the evolution of rocky planets into exo-Earths or exo-Venus, observations of rocky exoplanets orbiting in the overlap between the Habitable Zone (HZ) and the Venus Zone (VZ) are critical for expanding our understanding of the \textit{limits} of habitability. 
They occupy a regime in which atmospheric warming, cloud feedbacks, water-vapor dynamics, atmospheric escape, and planetary history interact to determine whether surface liquid water can persist. Studying them can reveal why their outcomes differ. Crucially, while the transition is arguably not governed by irradiation alone -- planets receiving similar stellar energy may evolve differently because of differences in atmospheric composition, rotation, initial water inventory, geological activity, and stellar ultraviolet radiation \citep[e.g.,][]{Watson1981, 
Kasting1983, Kasting1988, Hamano2013, Wordsworth2013, Kasting2015, Kodama2015, Way2016, Sossi2020, Hoening2021, Krissansen2021, Miyazaki2021, Turbet2021, Baumeister2023, Selsis2023, Constantinou2024, Constantinou2026, Raviprakash2026, White-Gianella2026} -- observing planets in this intermediate-irradiation range provides a natural laboratory for identifying the influence of stellar irradiation on water retention and on planetary habitability.

These observations will also complement upcoming flagship missions to Venus like VERITAS \citep{Smrekar2022}, DaVinci \citep{Garvin2022}, and EnVision \citep{ghail2017} as well as smaller missions like Venus Life Finder \citep{Seager2022} because today's Venus does not provide a direct record of its early history. Different combinations of initial conditions and evolutionary mechanisms may produce similar present-day states, but observing exoplanets with irradiation similar to a Young Venus can illuminate early Venus evolutionary pathways.

The HZ is defined as the orbital range around one or multiple stars where liquid water could be stable on an Earth-like rocky planet's surface and can be expressed as a function of the host star's temperature for specific atmospheric compositions \citep[e.g.,][]{Hart1979, Kasting1993, Abe2011, Kopparapu2013, Kaltenegger2013, Kaltenegger2013binary2, Cullum2014, Ramirez2014, Ramirez2016, Ramirez2017}. 
This region is currently the focus of possible atmospheric biosignature searches \citep[see e.g.,][]{Kaltenegger2017, Fujii2018, Schwieterman2018, Lichtenberg2024}. 
The limits of the empirical HZ \citep[e.g.,][]{Kasting1993, Kopparapu2013} are based on observations in our own Solar System: the inner \textit{Recent Venus} (RV) limit and the outer \textit{Early Mars} (EZ) limit, representing the flux at the time neither planet is thought to have had liquid surface water. However, the inner limit of the HZ is not well established because Venus' surface is young. 

A second important orbital distance range, closer to the star, is the Venus Zone (VZ) \citep{Kane2014} defined as the region that is expected to host Venus-like planets: hot, CO$_2$-dominated, and uninhabitable worlds \citep[e.g.,][]{Walker1975, Kasting1988, Yang2013, Way2016, Kopparapu2017, Way2020, Kaltenegger2023, Ostberg2023b}. The inner limit of the VZ is set to 25 times modern Earth's irradiation ($S_0$) in the definition, the insolation flux limit where Venus is thought to begin losing most of its atmosphere \citep{Zahnle_2017}. The outer limit is set to the \textit{Runaway Greenhouse} (RG) limit of the 1D atmospheric model, which also defines the Empirical HZ. 

\textbf{\textit{In a simplified approximation}}, the increasing surface temperature of planets closer to their star is thought to increase ocean evaporation, thereby increasing atmospheric water vapor concentration and providing a \textbf{\textit{mainly}} positive feedback that further increases the greenhouse effect. Note that the cloud feedback on surface temperature change is still being explored and depends strongly on assumptions about cloud properties such as particle size, density, elevation, and vertical extent, and modifies this simplified picture \citep[e.g.,][]{Wordsworth2013, Leconte2013, Selsis2023, Fan2024}. Generally, the increasing surface temperature of planets is projected to lead to a runaway greenhouse effect that can evaporate the whole ocean and dramatically increase the surface temperature by several hundred Kelvin \citep[for details on the climate feedback see e.g., ][]{Komabayasi1967, Ingersoll1969, Kasting1988, Goldblatt2012, Hamano2013}. 
In addition, alternative 3D global climate models suggest that solar irradiation above 0.95 $S_0$ for a young Venus (modeled as a slow 5,833-hour rotator) and 0.92 $S_0$ for a young Earth (modeled as a faster 24-hours rotator) would keep water from condensing on the surface \citep{Turbet2021}, initiating catastrophic water loss \citep[see e.g.,][]{Leconte2013, Kasting2014, Hamano2015, Wolf2015} at a much faster timescale than evaporation. 

Until the launch of JWST, this question remained critical — but unobservable. However, whether water initially condenses on such planets or not, what young Venus was like, and which of these transition planets are Earth- and which are Venus-like can now be explored through observations: e.g., models of transmission spectra for the transiting Super-Earth LP 890-9c on the edge of the inner HZ around an M6 star \citep{DelRez2022} differ considerably between a hot planet, a steamy planet caught in a runaway greenhouse state, and an exo-Venus \citep[e.g.,][]{Kaltenegger2023}. The changing compositions of the atmospheres in these different environments, which could arguably reflect evolutionary stages of hot rocky worlds, significantly affect their spectra \citep[e.g.,][]{Quirino2023, Gomez-Barrientos2023, Kaltenegger2023, Ostberg2023a}, thereby differentiating observations of exo-Venus' from those of hot exo-Earths.   

To identify the best targets in the region where we expect a change from Earth-like to Venus-like planets, we define the \textit{Transition Zone} (TZ) as the region where planets orbit in the overlap of the outer Venus Zone and the inner empirical Habitable Zone -- thus, where rocky exoplanets should transition from Earth-like to Venus-like conditions (see Discussion). 

However, no target list of rocky exoplanets in the TZ exists yet: 
In this paper, we identify 23 rocky exoplanets in the Transition Zone as targets that should transition from water-rich, Earth-like worlds to inhospitable, Venus-like worlds, and 2 rocky exoplanets in the Venus Evolution Zone with irradiation between that of a Young Venus and a modern Venus.
We analyze the demographics of rocky TZ and EV exoplanets and discuss follow-up observations that could provide critical new insights into the evolution of hot terrestrial planets and the limits of habitability. 

\section{Methods}
We use the exoplanet data from the \href{https://exoplanetarchive.ipac.caltech.edu/}{NASA Exoplanet Archive, Sep 2026} (NEA) \citep{Christiansen2025,NASAExoplanetArchive} and, if available, update the stellar parameters with \href{https://www.cosmos.esa.int/web/gaia/data-release-3}{\textit{Gaia}'s Data Release 3} (DR3) data \citep{gaiadr3,gaia2016} for stars with a renormalized unit weight error (RUWE) $\leq$ 1.4 (as suggested in the \href{https://gea.esac.esa.int/archive/documentation/GDR2/Gaia_archive/chap_datamodel/sec_dm_main_tables/ssec_dm_ruwe.html}{\textit{Gaia} Data Release Documentation}) to create a uniform sample. 
We translate changes in \textit{Gaia} DR3 stellar data into changes in planetary parameters because planetary parameters are derived from stellar ones, meaning a 10\% increase in stellar radius translates to a 10\% increase in planetary radius for transiting planets. If planets are missing a semi-major axis, we calculate it from stellar mass and orbital period; if only upper and lower bounds are given for eccentricity, we set it to zero in our calculations.

We first identify exoplanets in the TZ based on the average incident stellar irradiation at the planet's orbit, $S$, accounting for orbital eccentricity \citep{Bolmont2016}. These limits are set by the \textit{Recent Venus} limit on the inner edge and the \textit{Runaway Greenhouse} limit on the outer edge and can be expressed by a polynomial fit for main-sequence stars with temperatures between 2,300K and 10,000K \citep[e.g.,][]{Kopparapu2013} using Equations \ref{eq:irradiation}, \ref{eq:eccentricity}, and \ref{eq:limits}:

\begin{equation}
\label{eq:irradiation}
    \frac{S}{S_0} = \frac{L/L_\odot}{(a_{eff}/a_\oplus)^2}
\end{equation}
\begin{equation}
\label{eq:eccentricity}
    a_{eff} = a(1 - e^2)^\frac{1}{4}
\end{equation}
\begin{equation}
    \label{eq:limits}
    S_{\text{eff}} = S_{\text{eff}\odot} + A(T_*) + B(T_*)^2 + C(T_*)^3 + D(T_*)^4,
\end{equation}
where $L$ is the luminosity of the star, $a$ is the semi-major axis and $e$ is the eccentricity of the orbit. $T$ is the star's effective temperature in Kelvin, $S_0$ is modern Earth's flux, $T_* = T - 5780$ K,  and $S_{\text{eff}}$ is the effective solar flux at the inner and outer TZ limits in units of modern Earth flux (see Fig.1). 
The constants for Equation \ref{eq:limits} are given here for the Inner TZ Limit ($S_{\text{eff}\odot}$: 1.7753, A: $1.4316 \times 10^{-4}$, B: $2.9875 \times 10^{-9}$, C: $-7.5702 \times 10^{-12}$, D: $-1.1635 \times 10^{-15}$), 
and for the outer TZ Limit ($S_{\text{eff}\odot}$: 1.0512, A: $1.3242\times 10^{-4}$, B: $1.5418\times10^{-8}$, C: $-7.9895 \times {10^{-12}}$, D: $-1.8328 \times 10^{-15}$).

To address the question of which planets receive stellar irradiation between the Recent Venus level and modern Venus, we modify the Recent Venus limit by changing the $S_{\text{eff}}$ coefficient to 1.9291 to create a \textit{Modern Venus, MV} limit (see Fig.1). 
We identify planets in the \textit{Evolving Venus (EV)} region between the Recent Venus (RV) and Modern Venus (MV) flux analogs as the best targets for observing modern Venus equivalents. The constants for Equation \ref{eq:limits} at the Inner EV Limit are ($S_{\text{eff}\odot}$: 1.9291, A: $1.4316 \times 10^{-4}$, B: $2.9875 \times 10^{-9}$, C: $-7.5702 \times 10^{-12}$, D: $-1.1635 \times 10^{-15}$). The outer EV limit is the same as the inner TZ limit given above.

Planets in multiple-star systems are flagged in the NEA and also in our target list (see zenodo). However, none of the final rocky-planet targets in the TZ and EV are flagged as orbiting multiple stars. For exoplanets in multiple-star systems, the limits of the HZ and the TZ should be calculated slightly differently \citep[see][for how to account for the effect of multiple host stars on the flux at a planet's orbit]{Kaltenegger2013, Kaltenegger2013binary2, KaneHinkel2013}.

To identify potentially rocky exoplanets, we adopt a maximum radius of 2 $R_\oplus$ \citep[see e.g.][]{hill2023, bohl2026}; If exoplanets do not have measured radii, we adopt a minimum mass of $\leq$ $5\,M_{\oplus}$ \citep[see also][]{hill2023, bohl2026}. We denote them as diamonds in Figure \ref{fig:tvir_Zoom} for clarity because, without a known inclination of their orbital plane to our line of sight, their mass estimates only provide minimum mass constraints, which are less characteristic for a rocky planet than a 2$R_{\Earth}$ radius cutoff for transiting planets \citep[see e.g.,][]{Kaltenegger2017}. 

To support future observations, we calculate the transmission spectroscopy metric (TSM) \citep[following][]{Kempton2018} to guide transit observations and the maximum apparent separation ($\theta$) in combination with the star-planet contrast ratio ($F_p/F_*$) to guide target selection for direct imaging and secondary eclipse observations for each planet using Equations \ref{eq:tsm}, \ref{eq:contrast}, and \ref{eq:angularsep}. We used a TSM scale factor of 0.167 for all rocky planets as applied for the small temperate sample with insolation values between 0.2 and 2.0 times the Earth’s insolation and Rp $<$ 2.0 $R_{Earth}$ and scale factors binned by planet radii for non-rocky planets \citep[][]{Kempton2018}.
If these calculations require missing planetary mass or radius, we apply the mass-radius relation to determine TSM and contrast ratio for these calculations only.
\begin{equation}
    \label{eq:tsm}
    TSM = \beta (R_p) \left(\frac{R_p^3\ T_{eq}}{M_p\ R_*^2} \right) * 10^{-m_J/5}
\end{equation}
\begin{equation}
    \label{eq:contrast}
    F_p/F_* = \alpha \left(\frac{R_p}{a}\right)^2
\end{equation}
\begin{equation}
    \label{eq:angularsep}
    \theta = a (1+e)/d
\end{equation}
where $a$, $R_p$, and $M_p$ are the planet's semi-major axis, radius, and mass, respectively. $R_*$ and $m_J$ are the radius and J-magnitude of the host star. $\beta$ is a normalization constant used to calculate the TSM that depends on the planet's radius. $\alpha$ is the assumed planetary albedo; we model the planet as a Lambertian sphere with a Venus-like albedo of 0.7. Note that the contrast ratio can easily be scaled for other albedos, like Earth's albedo of 0.3. $\theta$ is the maximum apparent angular separation between the star and the planet, and \textit{d} is the distance to the system.

\section{Results}

To understand the emergence and decline of habitability on rocky worlds, and the shift from water-rich, Earth-like worlds to dry, Venus-like worlds, we identified exoplanets that receive flux between Earth's and Venus's: 23 rocky exoplanets in the Transition Zone, the overlap between the outer Venus Zone and the inner Habitable Zone, and 2 rocky planets in the Evolving Venus Zone.

\begin{figure*}
     \centering
     \includegraphics[width=0.9\linewidth]{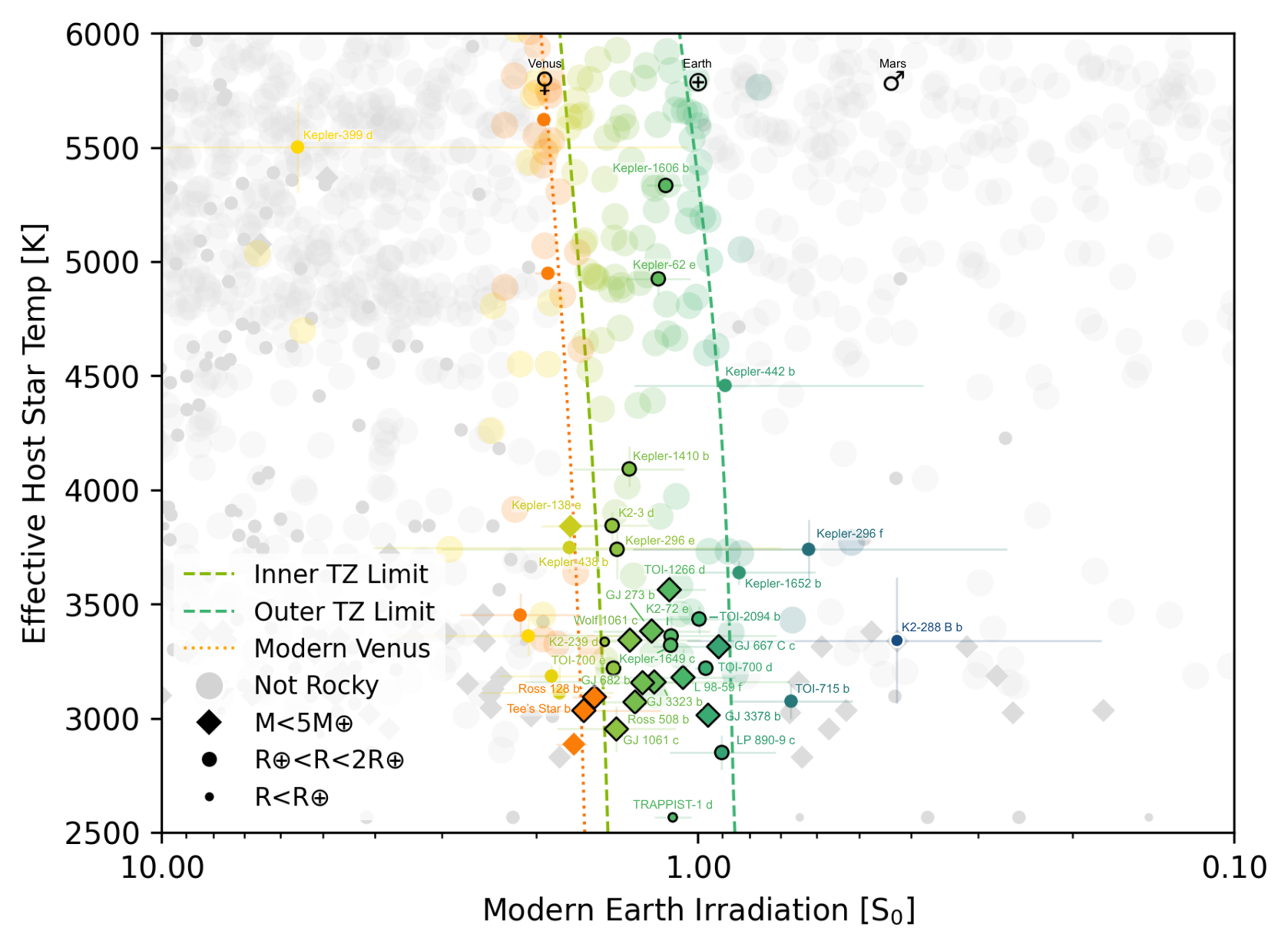}
    \caption{Temperature vs. incident irradiation in modern Earth's Irradiation ($S_0$) for the 23 rocky exoplanets in the Transition Zone (TZ) and the 2 rocky exoplanets in the Venus Evolution Zone (VE). The TZ limits and the modern Venus-flux equivalent are shown as dashed lines. The relative sizes of the symbols reflect the radius and mass of each planet. Transiting planets (circles) and planets with unknown radii (diamonds) are drawn solid for radii up to $ 2R _ {\ Earth} $ and if radii are unknown, minimum masses up to 5$M_{\Earth}$. Planets in the TZ are colored green, and in the EV, orange. 13 additional rocky planets lie outside the TZ with nominal values but have error bars that overlap the TZ, and 21 additional rocky exoplanets have error bars that overlap the EV zone; these are plotted without black outlines. Planets that are only in the EV are colored orange; planets whose uncertainties overlap with both regions default to green.}
    \label{fig:tvir_Zoom}
 \end{figure*}

We analyzed exoplanet data from the \href{https://exoplanetarchive.ipac.caltech.edu/}{NASA Exoplanet Archive} and updated stellar parameters with \textit{Gaia} DR3 data, if available, for stars with a renormalized unit weight error (RUWE) $\leq$ 1.4 \citep{gaiadr3} to provide a uniform sample. Specifically, the radii of 218 exoplanets (out of 4181 transiting planets with non-null stellar radii in NEA and Gaia DR3) increased above 2$R_{\Earth}$ due to these updates, while the radii of another 48 exoplanets decreased below that threshold. Here we focused on a specific stellar insolation range and identified 107 planets in the Transition Zone, among them 23 rocky exoplanets as well as 12 planets in the Evolving Venus Zone, including 2 rocky exoplanets (see Fig.\ref{fig:tvir_Zoom} and Table \ref{tab:TZ}) (see zenodo). 

Table \ref{tab:TZ} provides a sample list of rocky TZ and EV exoplanets and their host stars (full table available as supplement and on zenodo when accepted; see CSV files): Planet data (name, radius, and mass), incident stellar flux (Flux, Min, Max), flux at limits of the TZ (the empirical Recent Venus (RV) and Runaway Greenhouse (RG) limits). Note that stellar data are provided as given in the original source, NEA or Gaia DR3, while shown with only one or two significant figures in the sample table. All fluxes are provided in units of modern Earth flux ($S_0$). We also list the distance to the system in pc (d), the stellar temperature ($T_*$), the calculated maximum apparent angular separation ($\theta$), and planet-star contrast ratio ($F_p/F_*$). Planets in Table \ref{tab:TZ} are sorted by descending TSM values for those that transit (above the dotted row) to prioritize potential candidates for transmission spectroscopy, and by angular separation for direct imaging of those that do not transit (below the dotted row) for planets with ($F_p/F_*$) above $10^{-8}$ to prioritize potential candidates for direct imaging.

The top 5 candidates for both transit and non-transiting planets that are shown in the table \ref{tab:TZ} have stellar irradiantion ranging from 0.9 to 1.5 modern Earth's (transiting) and 0.9 to 1.35 modern Earth's (non-transiting), which allows starting to test whether significant transitions occure over this range of incident irradiation: LP 890-9c receives less flux than modern Earth, TOI-2094 b similar amounts, and TRAPPIST-1d, K2-239d and K2-3d more irradiation than moden Earth. For non-transiting planets, GJ 667C c and GJ 3378b receive less irradiation, and GJ 237 b, GJ 682b and Wolf 1061c receive more irradiation than modern Earth. Note that the irradiation from a redder star warms the surface of an Earth-like rocky planet more effectively than from a Sun-analog \citep[e.g.,][]{Kasting1993}.

Two planets on our list have been studied in more detail and observed with JWST already TRAPPIST-1d, and LP 890-9c. TRAPPIST-1d is the highest-ranked target for transmission spectroscopy observations on our list. The JWST observations rule out clear, hydrogen-dominated atmospheres, analogous to a haze-free Titan, cloud-free Venus, early Mars, or a cloud-free modern Earth \citep[e.g.,][]{Piaulet2025}. Thus, any atmosphere is likely thin or contains clouds or high-altitude aerosols; note that 3D global climate models predict water cloud formation at the terminator. Alternatively, if TRAPPIST-1 d is airless, evolutionary models indicate it must have formed with no more than 4 Earth oceans of water. The TRAPPIST-1 system is an important case study in the debate over whether M-star planets can retain their atmospheres during a young M-star's high-activity phase. However, even if the TRAPPIST-1 system planets were to be stripped of atmospheres, further observations of other M-star systems would be needed to confirm whether this is a standard evolution for M-star systems or not \citep[see e.g.,][and references therein]{Kaltenegger2017, Lichtenberg2024}.

The target in the TZ with the next highest TSM value is LP 890-9 c \citep{DelRez2022}, also orbiting a cool, red star, which could provide a range of possible environments, depending on initial water content and evolution between a hot Earth and a Venus-analog \citep[e.g.,][]{Kaltenegger2023, Quirino2023, Gomez-Barrientos2023}. LP 890-9 c was just observed in cycle 4 with JWST, which should provide insight into its environment once the data are published. K2-239d, with a similar TMS to LP 890-9 c has not yet been observed with JWST and would provide a very hot rocky world with an incident irradiation of 1.49 $S_0$ and a large radius of 1.9 Earth radii.

For the non-transiting planets, Wolf 1061c, GJ 667C c, and GJ 682 b, no spectral observations exist yet; however, the contrast ratios between 7.5 $10^{-7}$ and 4.4 $10^{-7}$ and angular separations between 23 and 17 mas make them interesting targets for direct imaging once the next generation of telescopes becomes available.

Figure \ref{fig:tvir_Zoom} shows the rocky exoplanets in the TZ and EV, the TZ and EV limits, the incident nominal stellar flux, and its uncertainty error bars. The relative sizes of the points reflect each planet's size. Circles denote transiting planets, while diamonds indicate planets with unknown radii. Planets in the TZ are shown as dark green, and in the EV as orange. 
Planets larger than 2$R_{\Earth}$ or planets more massive than a minimum mass of 5$M_{\Earth}$ with unknown radii are shown as faded shapes in all figures. 

13 additional planets lie outside the TZ with nominal values but have error bars that partially overlap the TZ; 21 have error bars that overlap the EV zone. These planets are all plotted without black outlines. Planets that are only in the EV are colored orange, and planets whose uncertainties overlap with both regions default to green \citep[see also][]{bohl2026, kane2014starHZ, kaltenegger2011}. We use “nominal” to refer to the value listed in the curated NEA table, as well as the curated \textit{Gaia} DR3 values if the stellar parameters were updated. This simplified approach also identifies which exoplanet host stars could benefit from additional observations to further constrain these errors (see Discussion) (zenodo link).

\begin{table*}[htbp]
\resizebox{\textwidth}{!}{
    \centering 
    \footnotesize
    \begin{tabular}{lclcccccrcclcc} 
        \toprule
        {Planet Name}& {Radius} & {Mass} & {Flux} & {Min} & {Max} & {$TZ_{in}$} & {$TZ_{out}$}  & {TSM} & {$\theta$} & {Contrast}  & {Age} & {$d$} & {${T}_{eff}$}\\
        & {R$_\oplus$} & {M$_\oplus$} & {S$_\oplus$} & {S$_\oplus$} & {S$_\oplus$} & {S$_\oplus$} & {S$_\oplus$}  & {} & {mas} & {} & {Gyr} & {pc} & {K}\\
        \midrule
        Transition Zone \\
        \midrule
        TRAPPIST-1 d & $0.78^{+0.1}_{-0.1}$ & $0.39^{\pm0.01}$ & 1.11 & 1.03 & 1.20 & 1.47 & 0.85 & 22.7 & 1.79 & 1.59e-6 & $7.6^{\pm2.2}$ & $12.47^{\pm0.01}$ & $2566^{\pm26}$ \\
        LP 890-9 c & $1.37^{+0.05}_{-0.04}$ &  & 0.90 & 0.72 & 1.13 & 1.49 & 0.86 & 6.92 & 1.23 & 1.49e-6 & $7.2^{+2.2}_{-3.1}$ & $32.43^{+0.07}_{-0.07}$ & $2850^{\pm75}$ \\
        K2-239 d & $0.77^{\pm0.07}$ &  & 1.49 & 1.42 & 1.63 & 1.51 & 0.87 & 6.82 & 2.21 & 1.6e-7 &  & $30.93^{+0.76}_{-0.08}$ & $3334.9^{+4.4}_{-4.8}$ \\
        TOI-2094 b & $1.9^{\pm0.1}$ &  & 0.99 & 0.72 & 1.37 & 1.52 & 0.87 & 3.50 & 2.72 & 2.5e-7 &  & $50.02^{+0.08}_{-0.08}$ & $3435^{+55}_{-65}$ \\
        K2-3 d & $1.46^{+0.05}_{-0.05}$ &  & 1.44 & 1.23 & 1.73 & 1.55 & 0.88 & 2.54 & 4.97 & 6.65e-8 & $3.7^{\pm1.0}$ & $44.07^{\pm0.11}$ & $3844^{+61}_{-63}$ \\
        $\cdots$ & $\cdots$ & $\cdots$ & $\cdots$ & $\cdots$ & $\cdots$ & $\cdots$ & $\cdots$  & $\cdots$ & $\cdots$ & $\cdots$  & $\cdots$ & $\cdots$ & $\cdots$ \\
        Wolf 1061 c &  & $3.4^{+0.4}_{-0.4}$ & 1.34 & 0.98 & 1.83 & 1.51 & 0.87 & 46.2 & 22.9 & 4.43e-7 & $6.4^{\pm2.5}$ & $4.306^{\pm0.001}$ & $3342^{\pm49}$ \\
        GJ 667 C c &  & $3.8^{+1.5}_{-1.2}$ & 0.914 & 0.723 & 1.21 & 1.51 & 0.87 & 21.5 & 17.6 & 2.6e-7 & 2 & $7.243^{+0.018}_{-0.009}$ & $3313.4^{+36.1}_{-2.4}$ \\
        GJ 682 b &  & $4.4^{+3.7}_{-2.4}$ & 1.3 & 0.91 & 1.5 & 1.50 & 0.86 & 44 & 17 & 7.4e-7 & $6.4^{\pm4.3}$ & $5.003^{+0.004}_{-0.003}$ & $3155.8^{+1.4}_{-0.5}$ \\
        GJ 273 b &  & $2.9^{+0.3}_{-0.3}$ & 1.22 & 0.94 & 1.56 & 1.52 & 0.87 & 48.7 & 16.923 & 3.48e-7 & $10.3^{\pm6.2}$ & $5.92^{\pm0.020}$ & $3382^{\pm49}$ \\
        GJ 3378 b &  & $2.3^{\pm0.4}$ & 0.96 & 0.94 & 0.98 & 1.49 & 0.86 & 12.2 & 12.52 & 2.4e-7 &  & $7.729^{+0.005}_{-0.008}$ & $3013.4^{+2.1}_{-0.4}$ \\
        \midrule
        Venus Evolution Zone \\
        \midrule
        Ross 128 b &  & $1.40^{\pm0.21}$ & 1.56 & 1.44 & 1.70 & 1.49 & 0.87 & 54.3 & 16.4 & 6.33e-7 & $7.28^{\pm3.29}$ & $3.3731^{\pm0.0007}$ & $3093.4^{+0.4}_{-0.4}$ \\
        Teegarden's Star b &  & $1.16^{+0.12}_{-0.11}$ & 1.63 & 1.17 & 2.25 & 1.495 & 0.87 & 76.5 & 6.96 & 2.09e-6 &  & $3.831^{+0.004}_{-0.004}$ & $3034^{\pm45}$ \\
        \bottomrule
    \end{tabular}
    }
    \caption{Sample table of the rocky Transit Zone (TZ) exoplanet data based on \textit{Gaia DR3} and NASA Exoplanet Archive (\textit{Gaia-NEA}). Planets are sorted by descending TSM values for those that transit (above the dotted row) and by angular separation ($\theta$) and contrast ratio for direct imaging of those that do not transit (below the dotted row). The two planets in the Venus Evolution Zone on the bottom receive stellar flux between that of a Young Venus and modern Venus.        
    Incident flux, nominal, minimum, and maximum values (Flux, Min, Max), and flux at the limits of the TZ are given in units of modern Earth flux ($S_0$). 
    The full table with a wider range of parameters is available in the supplement material and on zenodo (add link here when accepted).}
    \label{tab:TZ}
\end{table*}

\begin{figure}
    \centering
    \includegraphics[width=1\linewidth]{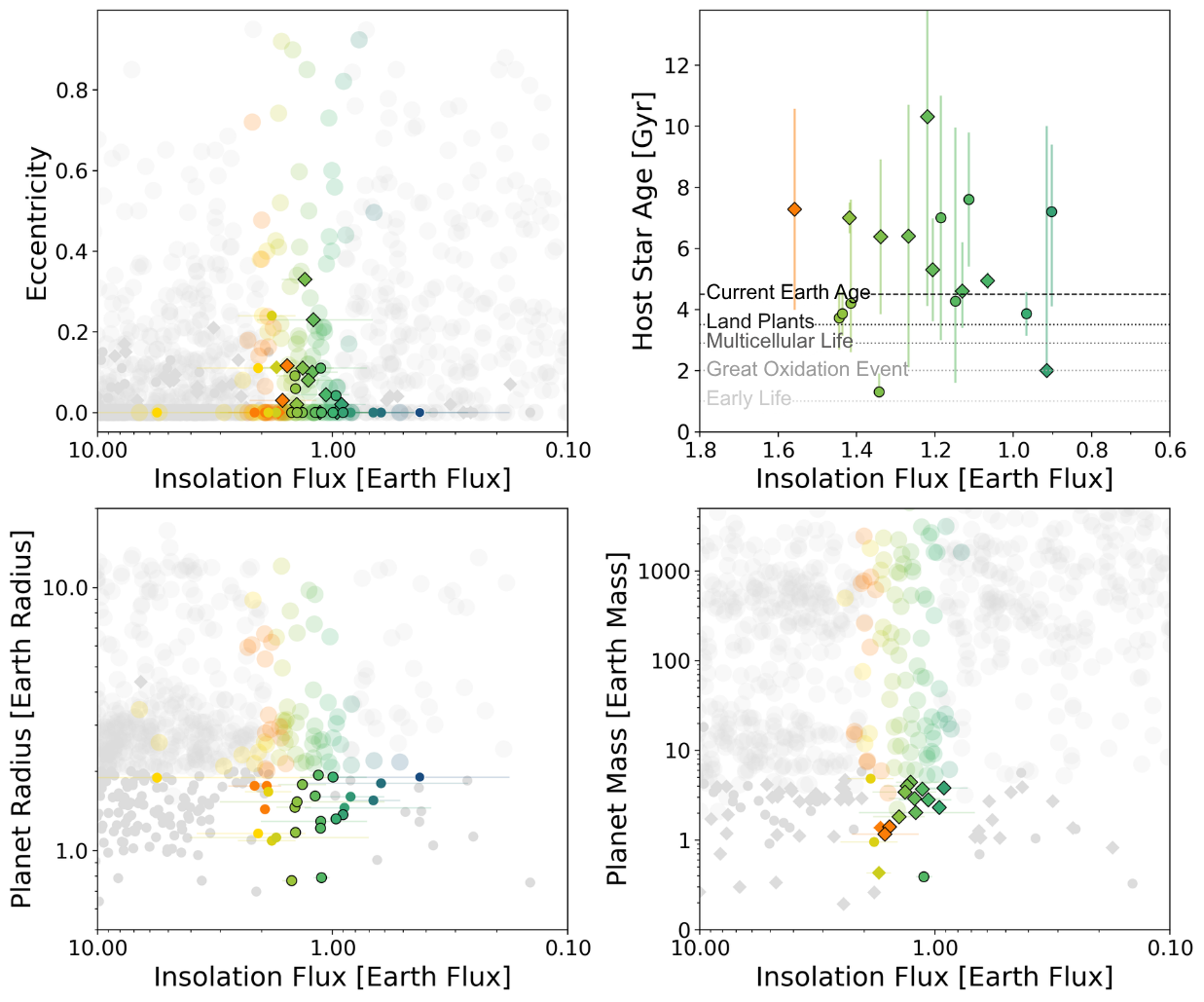}
    \caption{Characterization of rocky TZ planets compared to all known exoplanets: (top) eccentricity (left), and planetary minimum mass (right), and (bottom) radius (left), and age estimates for rocky TZ planets vs stellar irradiation. Planets whose nominal values do not place them in the TZ are shown without black outlines. 
    Exoplanets in the TZ are shown in shades of green. Rocky TZ planets are shown in solid color while non-rocky planets are semi-transparent. Note that we do not include eccentricities when only upper and lower bounds are given.}
    \label{fig:scatters}
\end{figure}

After identifying rocky planets in the TZ, we compare their demographics and characteristics with the overall exoplanet population. Figure \ref{fig:scatters} relates radius, mass, eccentricity, and stellar irradiation of the planets in the TZ (light to dark green) to that of all known exoplanets (black) and shows the age estimate for the TZ and EV systems, if available: (top) eccentricity vs incident stellar flux, and minimum mass versus incident irradiation and (bottom) radius versus incident irradiation and age estimates for the TZ systems, if available. 

Figure \ref{fig:scatters} shows that most rocky TZ planets have low to moderate eccentricity, with the two highest-eccentricity planets (Ross 508b, e=0.33, and GJ 3323 b, e=0.23) as interesting candidates to probe how eccentricity affects the stability of liquid water on a planet's surface. 

Because the TZ has fixed stellar-flux limits, rocky TZ planets occupy a small range on the insolation axis compared with overall exoplanet demographics in both the mass and radius panel. Due to their smaller size and mass, rocky TZ planets tend to be found more frequently around smaller host stars, where their signals are larger and thus easier to detect than those of similar planets orbiting more massive and larger host stars (see also Appendix A). As shown in Figure \ref{fig:tvir_Zoom} and Table \ref{tab:TZ}, our sample of TZ planets around hotter host stars was predominantly found by the Kepler mission.  

Of the 23 TZ planets, 17 planets have system age estimates, see Table \ref{tab:TZ} \citep[see][for details]{bohl2026}: GJ 1061 c, GJ 273 b, GJ 3323 b, GJ 667 C c, GJ 682 b, K2-3 d, Kepler-1410 b, Kepler-1606 b, Kepler-296 e, Kepler-62 e, L 98-59 f, LP 890-9 c, TOI-1266 d, TOI-700 d, TOI-700 e, TRAPPIST-1 d, Wolf 1061 c; 5 systems do not: GJ 3378 b, K2-239 d, K2-72 e, Kepler-1649 c, Ross 508 b, and TOI-2094 b. We provide context in Figure \ref{fig:scatters} for those age estimates by comparing them to important milestones in the evolution of life on Earth in the panel. Note that the exact times of earlier milestones are still debated on Earth, and the large uncertainties in these stellar age estimates are reflected in the large error bars in Table \ref{tab:TZ} and Figure \ref{fig:scatters}. 

Among the 17 TZ exoplanet host stars with age estimates, 9 stars have nominal ages older than Earth (GJ 273, TRAPPIST-1, LP 890-9, GJ 1061, Kepler-62, GJ 682, Wolf 1061, GJ 3323, L 98-59), one has a similar age to Earth (TOI-1266), and 6 stars have a younger age than Earth (Kepler-1606, Kepler-296, TOI-700, K2-3, GJ 667 C, Kepler-1410). Of the two planets in the Evolving Venus Zone, one has an age estimate (Ross 128), while the other does not (Teegarden's Star). Ross 128 has a nominal age older than Venus. Assuming nominal ages, this would allow a longer evolution time on 10 rocky TZ and 1 EV exoplanets than on Earth and Venus. 

These age estimates can help fill in the picture of how long different evolutionary stages from hot Earths to Venus could last: younger planets could still be hot Earths, while older ones could have already evolved into Venus analogs, though the timescale of water loss arguably depends on their initial water inventory, whether water condensed at all on a planet's surface, and climate- and cloud-feedback (see Discussion). These age estimates can also help interpret observations, e.g., if an exoplanet with a nominal age older than Earth still shows signs of water and temperate climate conditions, then that incident stellar irradiation can arguably be deemed safer against entering the runaway greenhouse stage than for a much younger planet that might still develop a runaway greenhouse state. 
While uncertainties in ages are extremely large (see \citep{bohl2026} for a discussion of age estimates), they can guide target selection, e.g., by identifying planets with similar incident irradiation but different age estimates to create a diverse age sample for comparison.

 \begin{figure}
     \centering
     \includegraphics[width=1\linewidth]{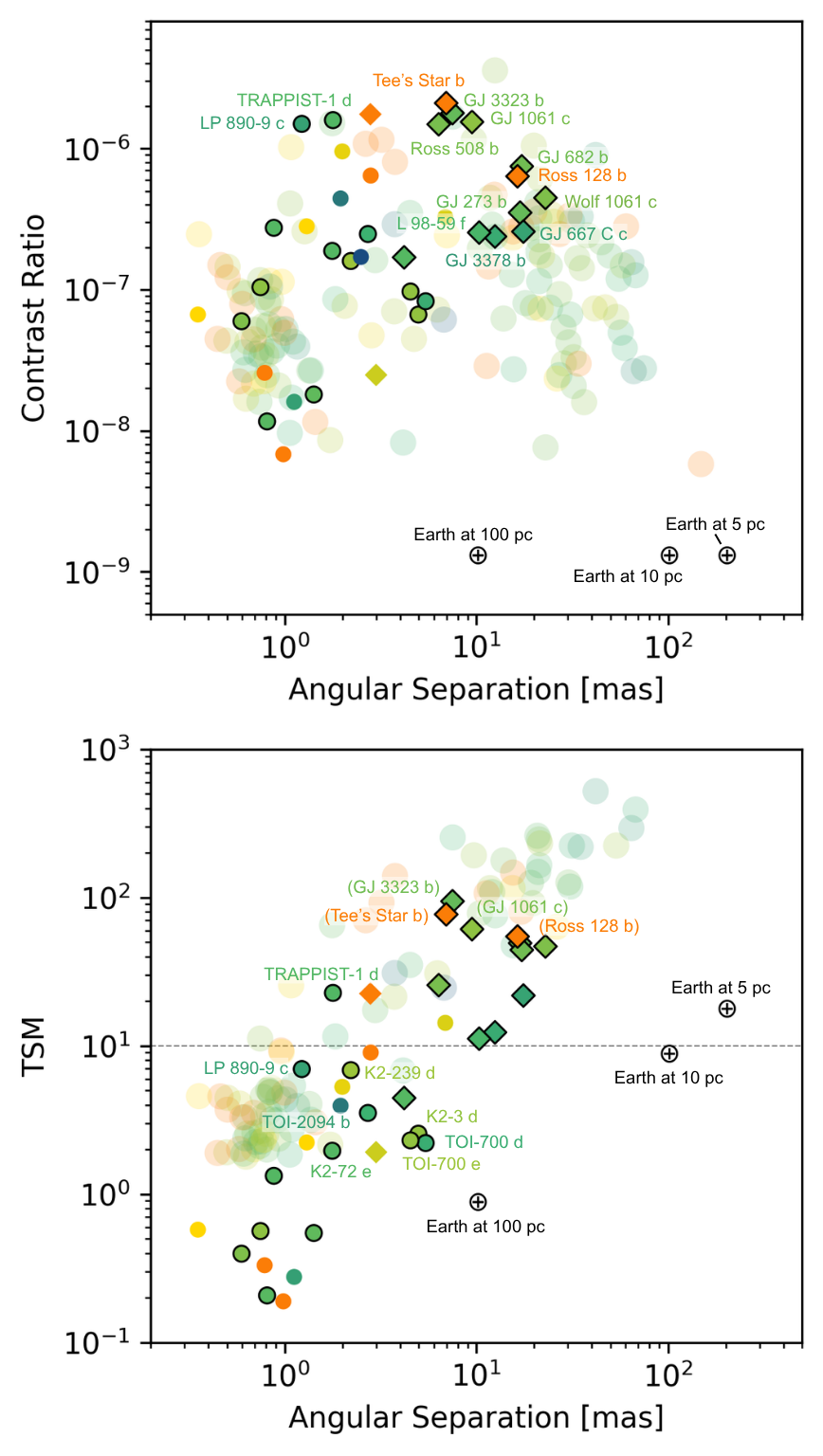}
     \caption{Apparent angular separation of rocky exoplanets in the Transition  Zone (TZ) and Venus Evolution Zone (VE) plotted versus contrast ratio between host star and planet (top) and versus transmission spectroscopy metric (TSM) (bottom). Planets whose nominal values place them in the TZ and EV are shown with black outlines, rocky planets in solid color, while non-rocky planets are semi-transparent. The colors are shades of green in the TZ and orange in the EV. For reference, we include a TSM of 10 as a dashed line and non-transiting RV planets for comparison only, indicated by the RV planet names in parentheses, and we take the Sun's J-magnitude from \citep{Willmer2018}. }
     \label{fig:tsm_contrast}
 \end{figure}

Ongoing transit observations with JWST use the TSM as a metric of observability and have observed both TRAPPIST-1d and LP 890-9c with a TSM of about 23 and 7, respectively. For lower TSM values and non-transiting planets, upcoming telescopes like the Extremely Large Telescope (ELT) and designs like NASA's Habitable World Observatory (HWO) and the European LIFE design are planned to enable transit observations and direct imaging of additional rocky planets in our target list \citep[e.g.,][]{Konrad2023, Kane2026}. Thus, Figure \ref{fig:tsm_contrast} shows the apparent angular separation of rocky exoplanets in the TZ versus the (top) host-star-to-planet contrast ratio and the (bottom) transmission spectroscopy metric (TSM). Rocky planets are plotted as opaque circles or diamonds and non-rocky planets are plotted as larger, semitransparent circles. RV planets that do not transit are shown in the TSM plot for reference only, with their names in brackets, e.g., as (Ross 128b). The colors correspond to the planet irradiation using the same convention as in Figure \ref{fig:tvir_Zoom}

Exoplanets in the TZ provide a critical observational opportunity to explore what influences whether a planet can provide and maintain habitable conditions or not, how long rocky planets can hold on to their water, where the inner edge of the Habitable Zone lies, and how the limits of habitability change with planetary characteristics and stellar hosts. In addition, they provide a critical calibration set for interpreting spectral observations of rocky exoplanets. 

\section{Discussion}
Planets in the Transition Zone let us test whether and how irradiation influences habitability. While the inner limit of the habitable zone is unlikely to be a simple distance based on stellar irradiation because it depends on atmospheric physics, clouds, water availability, and the planet’s evolutionary history, stellar irradiation is a key component of a planet's environment. 
Studying planets in the TZ tests how stellar irradiation influences the habitability puzzle. Observing TZ planets can test three competing possibilities: i) that there is a gradual transition where increasing stellar irradiation progressively warms the climate and increases water loss, ii) that there is a threshold, a critical point of irradiation that triggers rapid irreversible water loss, and iii) that there are multiple stable outcomes and planets receiving similar irradiation can end up in very different climate states because their clouds, atmospheric circulation, rotation, or initial conditions differ.

\subsection{Stellar Data Updates Gaia DR3} 
Stellar data are critical for assessing which planets are in the TZ and EV because the limits -- like the HZ limits -- depend on incident stellar irradiation, which, in turn, is modeled as a polynomial function of $T_*$ (see Methods). Updates to the host star's properties affect both the incident flux reaching the planet's orbit and, consequently, the system's and its planets' TZ orbital distances. 

In addition, planetary properties like a transiting planet's radius depend on the host star's parameters because we calculate planetary radius and mass from the host star's properties. When we updated the stellar parameters of the exoplanet host stars in NEA with available data from \textit{Gaia} DR3 for stars with a renormalized unit weight error (RUWE) $\leq$ 1.4 \citep{gaiadr3}, the radii of 4179 transiting exoplanets (out of 6360 total planets) changed because of differences in their host stars' radii between NEA and Gaia DR3. For our analysis, 218 transiting exoplanets had radii that increased above the $2\,R_\oplus$ threshold, and 48 had radii that decreased below it due to GAIA DR3 stellar parameter updates to the NEA. Thus, some exoplanets gain—and many others lose—their assignment as rocky planets due to updated stellar data (see Table \ref{tab:tableNEAvsGaia}). 
17 rocky TZ planets overlap in both lists ( NEA-\textit{Gaia}: 23, NEA-only: 21) and 1 in the EV Zone (NEA-\textit{Gaia}: 2, NEA-only: 3).
Looking in detail at the changes for the Transition Zone we find that 6 planets enter the rocky planet TZ target list when updating NEA-Only stellar data with \textit{Gaia}-DR3: GJ 667 C c, GJ 682 b, K2-239 d, Kepler-1606 b,  Kepler-1649 c, and TOI-700 d. 2 of these 6 planets are missing stellar radii in NEA (GJ 667 C c and GJ 682 b), 3 have different $T_{\text{eff}}$ and stellar radius (K2-239 d, Kepler-1649 c, and TOI-700 d), and 1 planet's radius changed from non-rocky to rocky due to the stellar radius update (Kepler-1606 b). 
In contrast, 4 planets no longer qualify as rocky TZ targets with updated \textit{Gaia}-DR3 stellar data: Kepler-1638 b, Kepler-440 b, Kepler-452 b, and Ross 128 b. All 4 of these planets have different $T_{\text{eff}}$ and stellar radii in NEA and \textit{Gaia}-DR3.  Note that Ross 128 b moved from the TZ to the EV with updated GAIA stellar data. The Appendix provides the NEA-only list of TZ exoplanets for comparison with literature research that uses NEA-only stellar parameters.

Looking in detail at the changes for the Evolving Venus Zone, we find that 1 planet overlaps in both lists: Teegarden's Star b. Ross 128 b is not part of the NEA-only EV dataset because of its different $T_{\text{eff}}$ and stellar radii in NEA versus \textit{Gaia}-DR3.
Updating the NEA stellar parameters to \textit{Gaia}-DR3 data removes Proxima Cen d and Kepler-69 c from the EV list.

While there is still a large overlap among rocky planets in the TZ (see Table \ref{tab:tableNEAvsGaia}) and some overlap in the EV Zone, this comparison shows how critical stellar parameters are for determining which planets orbit in the TZ and their planetary characteristics, since these are derived from stellar data. Further observations would be especially useful for constraining the stellar parameters of systems that differ between the lists \citep[see, e.g., initial work by][]{Laverny2025}. 

\begin{table}
    \centering 
    \begin{tabular}{ll} 
        \toprule
        Category & Planet Names \\
        \midrule
        \midrule
        Transition Zone \\
        \midrule
         \multirow{5}{*}{Overlap (15)}
         & GJ 1061 c, GJ 273 b, GJ 3323 b, \\ 
         & GJ 3378 b, K2-3 d, K2-72 e, \\ 
         & Kepler-1410 b, Kepler-296 e, \\
         & Kepler-62 e, L 98-59 f, LP 890-9 c, \\
         & Ross 508 b, TOI-1266 d, TOI-2094 b, \\
         & TOI-700 e, TRAPPIST-1 d, \\
         & Wolf 1061 c \\ 
        \midrule
         NEA-\textit{Gaia} (6)& GJ 667 C c, GJ 682 b, K2-239 d, \\ 
         not NEA-Only & Kepler-1606 b, Kepler-1649 c, \\
         & TOI-700 d \\ 
        \midrule
        NEA-Only (4)& Kepler-1638 b, Kepler-440 b, \\
        not NEA-\textit{Gaia} & Kepler-452 b*, Ross 128 b\\ 
        \midrule
        \midrule
        Venus Evolution Zone \\
        \midrule
        Overlap (1) & Teegarden's Star b \\
        \midrule    
        NEA-\textit{Gaia} (1) & Ross 128 b  \\
        not NEA-Only \\
        \midrule    
        NEA-Only (2) & Kepler-69 c$^\dagger$ \\
        not NEA-\textit{Gaia} & Proxima Cen d\\
        \bottomrule
    \end{tabular}
    \begin{tablenotes}
	\small {
	\item{* Indicates a controversial planet in the NEA}
    \item{$\dagger$ Indicates a planet that is in the VZ due to rounding effects.}}
    \end{tablenotes}
    \caption{Comparison of rocky planets in the Transition Zone and Venus Evolution Zone between NASA Exoplanet Archive only host star data (NEA-only) vs. updated stellar data based on {\it Gaia} DR3 stellar parameters (NEA-{\it Gaia}). 
    }
    \label{tab:tableNEAvsGaia}
\end{table}

Note that we do not include alternative limits of the VZ here that are still being discussed, such as a potential new Outer VZ Limit that spans most of the HZ \citep{Vidaurri2022} or potential changes to the inner limits of the HZ for specific planetary conditions \citep[e.g.,][]{Abe2011, Yang2014}, which would expand the number of planets in the overlapping TZ region. While new limits for the VZ and HZ are being discussed, the current analysis identifies core planets in the TZ.

\section{Conclusion}
Planets in the Transition Zone provide key insights into when a rocky planet changes from an exo-Earth to an exo-Venus, and whether a minimum irradiation is required to form oceans -- temporarily or permanently -- and to keep them. Observing these planets can explore water retention in the TZ and illuminate the limits of habitability.

We analyzed the NASA Exoplanet Archive, updated stellar parameters with \textit{Gaia} DR3 data when available, and identified 23 rocky exoplanets in the Earth-to-Venus Transition Zone and 2 rocky exoplanets in the Evolving Venus Zone. These rocky planets are critical targets because they offer the best exoplanet candidates for understanding the transition between habitable and non-habitable conditions. 

Our Transition Zone rocky planet target catalog serves as input for observers to identify targets that address fundamental questions about a rocky planet's evolution and the limits of habitability. We also quantified their potential for follow-up transit observations with telescopes such as JWST and ELTs, and for future direct imaging with telescopes such as ELT, HWO, and LIFE.  

While we cannot observe the development of Venus and Earth from formation to now in our own Solar System, rocky planets in the TZ can offer rare insights into the possible climate evolution of terrestrial planets, from hot exo-Earths to exo-Venus, and into the limits of habitability for rocky worlds on the edge.
They connect exoplanet observations to the unsolved Solar System puzzle of why Earth and Venus provide such contrasting outcomes. 

Characterizing exoplanets in the Transition Zone is critical to investigating the open questions of the emergence and decline of habitability on rocky worlds, the evolution from an exo-Earth to an exo-Venus, and Earth’s future.

\appendix

\section{Supplement information}
Table \ref{tab:TZ_NEA-only} provides a sample list of rocky TZ exoplanets and their host stars (full table available on zenodo when accepted) based on NEA-only data: Planet data (name, radius, and mass), incident stellar flux (Flux, Min, Max), and flux at the limits of the TZ (the empirical Recent Venus (RV) and Runaway Greenhouse (RG) limit). All fluxes are provided in units of modern Earth flux ($S_0$). We list the maximum apparent angular separation ($\theta$), planet-star contrast ratio ($F_p/F_*$), distance to the system in pc (d), and stellar temperature ($T_*$). Planets are sorted by descending TSM values for those that transit (above the dotted row) to prioritize potential candidates for transmission spectroscopy, and by angular separation for direct imaging of those that do not transit (below the dotted row) for planets with ($F_p/F_*$) above $10^{-8}$ to prioritize potential candidates for direct imaging.

\begin{table*}[htbp]
    \centering 
    \footnotesize 
\resizebox{\textwidth}{!}{
    \begin{tabular}{lclcccccrccccc} 
        \toprule
        {Planet Name}& {Radius} & {Mass} & {Flux} & {Min} & {Max} & {$TZ_{in}$} & {$TZ_{out}$}  & {TSM} & {$\theta$} & {Contrast}  & {Age} & {$d$} & {${T}_{\text{eff}}$}\\
        & {R$_\oplus$} & {M$_\oplus$} & {S$_\oplus$} & {S$_\oplus$} & {S$_\oplus$} & {S$_\oplus$} & {S$_\oplus$}  & {} & {mas} & {} & {Gyr} & {pc} & {K}\\
        \midrule
        TRAPPIST-1 d & $0.79^{+0.01}_{-0.01}$ & $0.39^{\pm0.01}$ & 1.11 & 1.03 & 1.20 & 1.47 & 0.85 & 22.7 & 1.79 & 1.59e-6 & $7.6^{\pm2.2}$ & $12.47^{\pm0.01}$ & $2566^{\pm26}$ \\
        LP 890-9 c & $1.37^{+0.05}_{-0.04}$ &  & 0.90 & 0.72 & 1.13 & 1.49 & 0.86 & 6.92 & 1.23 & 1.49e-6 & $7.2^{+2.2}_{-3.1}$ & $32.43^{+0.07}_{-0.07}$ & $2850^{\pm75}$ \\
        TOI-700 e & $0.95^{+0.09}_{-0.07}$ &  & 1.27 & 1.06 & 1.54 & 1.52 & 0.87 & 3.76 & 4.6 & 6.42e-8 & $3.86^{\pm0.72}$ & $31.12^{\pm0.02}$ & $3459^{\pm65}$ \\
        TOI-2094 b & $1.9^{\pm0.1}$ &  & 0.99 & 0.72 & 1.37 & 1.52 & 0.87 & 3.50 & 2.72 & 2.5e-7 &  & $50.02^{+0.08}_{-0.08}$ & $3435^{+55}_{-65}$ \\
        K2-3 d & $1.46^{+0.05}_{-0.05}$ &  & 1.44 & 1.23 & 1.73 & 1.55 & 0.88 & 2.54 & 4.99 & 6.65e-8 & $3.72^{\pm1.03}$ & $44.07^{\pm0.10}$ & $3844^{+61}_{-63}$ \\
        $\cdots$ & $\cdots$ & $\cdots$ & $\cdots$ & $\cdots$ & $\cdots$ & $\cdots$ & $\cdots$  & $\cdots$ & $\cdots$ & $\cdots$  & $\cdots$ & $\cdots$\\
        Wolf 1061 c &  & $3.41^{+0.43}_{-0.41}$ & 1.34 & 0.98 & 1.83 & 1.513 & 0.87 & 46.2 & 22.9 & 4.43e-7 & $6.38^{\pm2.53}$ & $4.306^{\pm0.001}$ & $3342^{\pm49}$ \\
        GJ 273 b &  & $2.89^{+0.27}_{-0.26}$ & 1.22 & 0.94 & 1.56 & 1.52 & 0.87 & 48.7 & 16.92 & 3.48e-7 & $10.31^{\pm6.20}$ & $5.92^{\pm0.02}$ & $3382^{\pm49}$ \\
        Ross 128 b &  & $1.40^{\pm0.21}$ & 1.47 & 1.17 & 1.87 & 1.50 & 0.87 & 64.2 & 16.4 & 6.33e-7 & $7.28^{\pm3.29}$ & $3.3745^{\pm0.0008}$ & $3192^{\pm60}$ \\
        GJ 3378 b &  & $2.3^{\pm0.4}$ & 0.90 & 0.8 & 1.05 & 1.51 & 0.87 & 19.2 & 12.53 & 2.4e-7 &  & $7.722^{\pm0.005}$ & $3340^{\pm60}$ \\
        L 98-59 f &  & $2.8^{\pm0.3}$ & 1.09 & 0.92 & 1.30 & 1.52 & 0.87 & 14.4 & 10.34 & 2.5e-7 & $4.94^{\pm0.28}$ & $10.619^{\pm0.003}$ & $3415^{\pm60}$ \\
        \bottomrule
    \end{tabular}
    }
    \caption{Sample table of the rocky Transit Zone (TZ) NEA-only exoplanet data. Planets are sorted by descending TSM values for those that transit (above dotted row), and by angular separation ($\theta$) and contrast ratio for direct imaging of those that do not transit (below dotted row). 
    Incident flux, nominal, minimum, and maximum values (Flux, Min, Max), flux at the limits of the TZ are given in units of modern Earth flux ($S_0$). 
    The full table with a wider range of parameters is available on zenodo (add link here when accepted).}
    \label{tab:TZ_NEA-only}
\end{table*}

\begin{figure}
    \centering
    \includegraphics[width=0.8\linewidth]{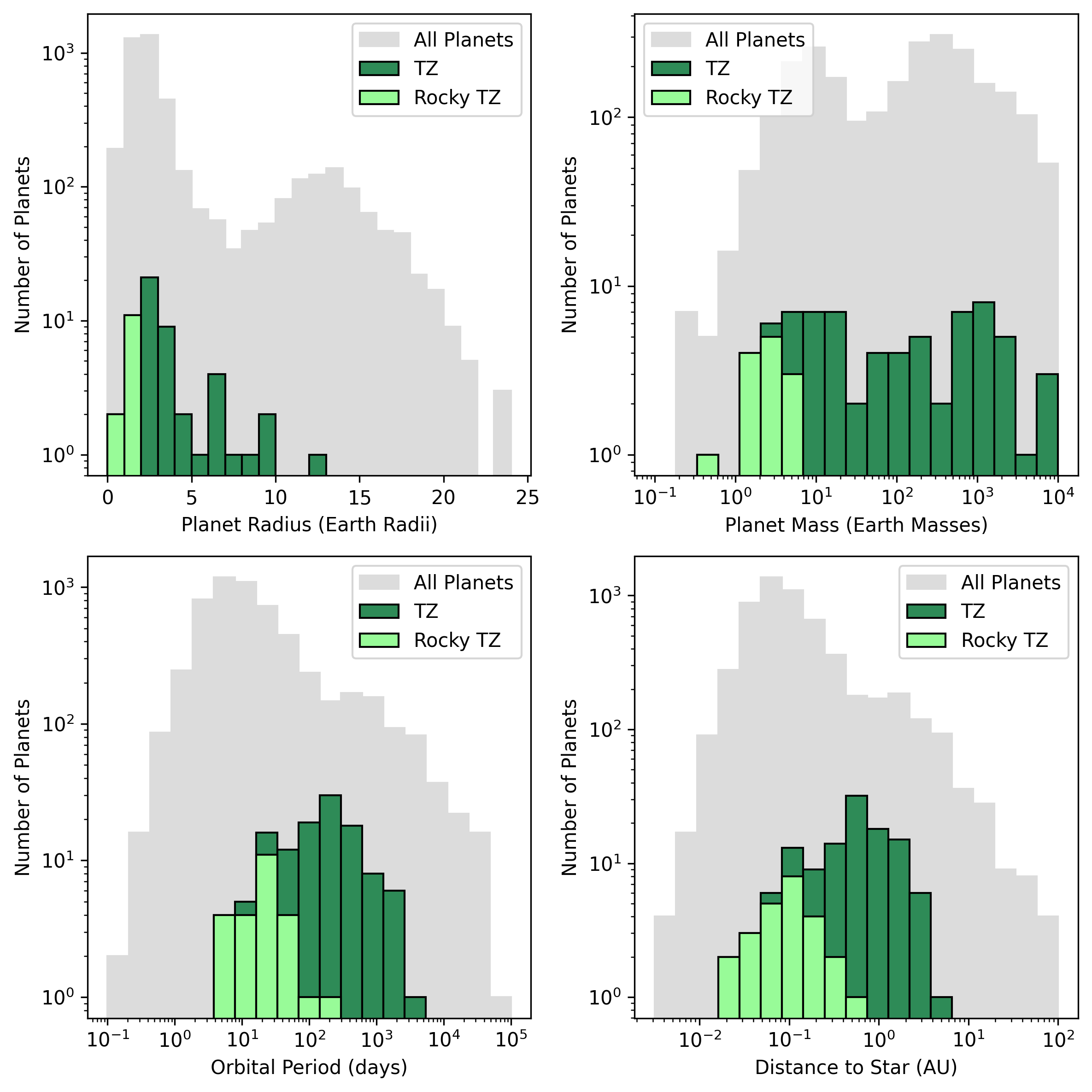}
    \caption{Demographics of radius, mass, orbital distance, and host star temperatures for exoplanets in the Transition Zone (dark green=rocky TZ planets, light green=all TZ planets) compared to all confirmed exoplanets (grey).}
    \label{fig:hists}
\end{figure}

Figure \ref{fig:hists} compares the demographics (radius, mass, orbital period, and orbital distance) of exoplanets in the Transition Zone (green=rocky TZ planets, yellow=all TZ planets) with all confirmed exoplanets (grey). The top two panels show the radius (top-left) and minimum mass (top-right) of the rocky TZ planets, subject to selection effects, compared with the overall exoplanet demographics. Note that the radius panel shows transiting planets, while the mass histogram shows RV planets, since those have observed radii and observed minimum masses that are not based on assumed mass-radius relations.

\clearpage
\bibliographystyle{aasjournal}
\bibliography{bib}
\end{document}